\documentclass[journal=jctcce,manuscript=article]{achemso}
\setkeys{acs}{usetitle=true}

\usepackage[version=3]{mhchem} 

\usepackage{amsmath,amssymb}
\usepackage{graphicx,color,subfigure}
\usepackage{amsfonts}
\usepackage{comment}
\usepackage{placeins}
\usepackage{fixltx2e}

\usepackage{sectsty}
\usepackage{balance}
\usepackage{xspace}
\usepackage{algorithm,algorithmic}

\usepackage{epsfig,amssymb,amsmath,amsthm,amsfonts,amsbsy,mathrsfs}
\usepackage{graphicx,color}
\usepackage{amsmath}
\usepackage{amssymb}
\usepackage{epstopdf}
\usepackage{threeparttable}
\usepackage{graphicx} 
\usepackage{lastpage}
\usepackage[format=plain,justification=raggedright,singlelinecheck=false,font=small,labelfont=bf,labelsep=space]{caption}

\allowdisplaybreaks

\usepackage[version=3]{mhchem} 

\newcommand{\br}{\mathbf{r}}
\newcommand{\bR}{\mathbf{R}}
\newcommand{\bk}{\mathbf{k}}

\title[Gradient]
{Perturbative variational quantum algorithms for material simulations}

\author{Jie Liu}
\email{liujie86@ustc.edu.cn} 
\affiliation{Hefei National Laboratory, University of Science and Technology of China, Hefei 230088, China}

\author{Zhenyu Li}
\email{zyli@ustc.edu.cn}
\affiliation{Hefei National Laboratory, University of Science and Technology of China, Hefei 230088, China}
\alsoaffiliation{Hefei National Research Center for Physical Sciences at the Microscale, University of Science and Technology of China, Hefei, Anhui 230026, China} 

\author{Jinlong Yang}
\email{jlyang@ustc.edu.cn}
\affiliation{Hefei National Laboratory, University of Science and Technology of China, Hefei 230088, China}
\alsoaffiliation{Hefei National Research Center for Physical Sciences at the Microscale, University of Science and Technology of China, Hefei, Anhui 230026, China}

\begin{document}

\begin{abstract}
{Reducing circuit depth is essential for implementing quantum simulations of electronic structure on near-term quantum devices. In this work, we propose a variational quantum eigensolver (VQE) based perturbation theory algorithm to accurately simulate electron correlation of periodic materials with shallow ansatz circuits, which are generated from Adaptive Derivative-Assembled Pseudo-Trotter or Qubit-Excitation-based VQE calculations using a loose convergence criteria. Here, the major part of the electron correlation is described using the VQE ansatz circuit and the remaining correlation energy is described by either multireference or similarity transformation-based perturbation theory. Numerical results demonstrate that the new algorithms are able to accurately describe electron correlation of the LiH crystal with only one circuit parameter, in contrast with $\sim$30 parameters required in the adaptive VQE to achieve the same accuracy. Meanwhile, for fixed-depth Ans\"atze, e.g. unitary coupled cluster, we demonstrate that the VQE-base perturbation theory provides an appealing scheme to improve their accuracy.}
\end{abstract}

\section{Introduction}
The exact solution of the Schr\"odinger equation on a classical computer is blocked by the exponential wall problem, that is, an exponential increase of computational complexity with increasing electron number. Quantum computing provides a new computational paradigm to solve the Schr\"odinger equation with favorable scaling~\cite{AspDutLov05,DuXuPen10,PerMcCSha14,MalBabKiv16,KanMezTem17,HemMaiRom18,AruAryBab20,Pre18,CaoRomOls19,McAEndAsp20}, which is critical for accelerating material design and drug discovery. However, in the current stage, quantum simulations of electronic structure remain a challenging task due to the presence of noise. Given limited qubit counts and circuit depth, variational quantum eigensolver (VQE)~\cite{PerMcCSha14,McCRomBab16,RomBabMcC18,AruAryBab20} is one of the most popular techniques to simulate electronic structure of molecular and periodic materials on a near-term quantum computer. While designing a good VQE ansatz to give consideration to both circuit depth and expressivity is still an open problem.   

Much effort has been devoted to building low-depth Ans\"atze for an appropriate description of electron correlations within the framework of the VQE~\cite{KanMezTem17,KivMcCWie18,BabWieMcC18,GriEcoBar19,LeeHugHea19,MatKur20}. For example, in contrast to the chemically inspired unitary coupled cluster (UCC) ansatz~\cite{PerMcCSha14,SheZhaZha17,RomBabMcC18,LeeHugHea19}, adaptive variational quantum algorithms, such as Adaptive Derivative-Assembled Pseudo-Trotter (ADAPT)~\cite{GriEcoBar19,LiuWanLi20} and iterative qubit coupled cluster VQE~\cite{RyaLanGen20}, were recently proposed to iteratively approach the exact eigenenergy of an electronic Hamiltonian. Furthermore, a hardware-heuristic ansatz that employs block-entangled circuit structures was suggested to implement quantum simulations of electronic structure on a noisy intermediate-scale quantum (NISQ) device~\cite{KanMezTem17,BarGonSok18,ZenFanLiu23}. {However, the circuit depth for applying these Ans\"atze to accurately treat electron correlation of complex systems is still prohibited on near-term quantum devices. To further reduce circuit depth often implies loss of accuracy in the VQE calculations. As such, it is necessary to develop post-VQE algorithms to restore the exact electron correlation energy for VQE calculations with shallow ansatz circuits.}

Previous studies of the ADAPT-VQE revealed that most of electron correlation could be captured by a shallow ansatz circuit with a few parameters~\cite{GriEcoBar19,LiuLiYan21}. { However, quantum circuits required to restore the rest of the electron correlation that is often attributed to the dynamic correlation are too deep to implement on NISQ devices~\cite{PerMcCSha14,GriEcoBar19,LeeHugHea19}. Note that this behaviour of slow convergence in the correlation energy is not unique to the ADAPT-VQE. In selected configuration interaction (SCI) calculations, a small fraction of the determinants of the full configuration interaction space is able to recover most of the correlation energy while a large number of additional determinants are necessary to predict the exact correlation energy~\cite{GarSceGin18}. 
Instead of increasing the circuit depth, perturbation theory (PT) is expected to be an alternative strategy to efficiently account for the dynamic correlation as commonly done in the quantum chemistry community.}

Perturbation theory has a long history in approximately solving the Schr\"odinger equation. Second-order M{\o}llet Plesset PT based on the Hartree-Fock reference state is a popular approach to treat the dynamic electron correlation in weakly correlated systems. Furthermore, combining multireference wave function methods, such as complete active space self-consistent field~\cite{SzaMulGid12}, density matrix renormalization group~\cite{SonCheMa20} and SCI~\cite{SmiMusHol17,GarSceGin18,ZhaLiuHof20,YaoUmr21}, with PT has been successfully applied to accurately describe electron correlation with moderate computational scaling~\cite{RooAndFul96,Pul11,LisNacAqu18}. Here, the static correlation is mainly described by a multiconfigurational wave function expanded in a small active space and the dynamic correlation beyond the active space is accounted for by PT~\cite{VogMaOls17}. 

Recently, integrating PT in quantum simulations of electronic structure has also attracted broad attention for an accurate and efficient description of electron correlation. Tammaro and coworkers proposed N-electron valence PT formulated in the VQE framework to study the relative stability of hydroxide anion and hydroxyl radical~\cite{TamGalRic23}. { Ryabinkin and coworkers established a posteriori PT correction based on the effective Hamiltonian in the qubit representation. Here, both of them focused on developing a new VQE-PT algorithm for molecular systems.} An efficient implementation of PT with no training or optimization process on a quantum computer was presented for treating a weak external field perturbation~\cite{LiJonKai23}, { in which the eigenenergy and corresponding eigenstates of the unperturbed Hamiltonian are assumed to be exactly known. In addition, a perturbative quantum dynamic simulation algorithm using the Dyson series expansion has been proposed for the solution of large quantum dynamics problems on NISQ hardware~\cite{SunEndLin22}. However, integrating the VQE with PT under periodic boundary condition for an accurate description of correlated solid materials is still lacking.}

{ In this work, we propose a new VQE-based PT algorithm for an accurate description of both static and dynamic electron correlations for periodic materials with shallow ansatz circuits { or to say ansatz circuits with few parameters in the context of the adaptive VQE. Here, the ADAPT-VQE calculations are first performed using a loose convergence criteria to generate low-depth circuits.} After that, two approaches, named VQE-based multireference PT (VQE-MRPT) and VQE-based similarity-transformed PT (VQE-STPT), are formulated to implement the VQE-PT algorithm for periodic systems based on the K2G scheme, which folds Bloch Hartree-Fock orbitals sampled at a set of $k$ points into supercell Hartree-Fock orbitals at $\Gamma$ point~\cite{LiuWanLi20}.} In the VQE-MRPT approach, a multiconfigurational reference wave function is prepared using the ADAPT ansatz~\cite{GriEcoBar19} or the UCC with single and double excitations (UCCSD) ansatz. Then, a set of anti-Hermitian operators are applied to this reference wave function to build the first-order interacting subspace that consists of a set of multiconfigurational basis functions orthogonal to the reference state. In the VQE-STPT scheme, the reference wave function is the Hartree-Fock state and an effective Hamiltonian is build by using a unitary operator, namely the ADAPT ansatz, to perform the similarity transformation of the electronic Hamiltonian. { The VQE-PT algorithms are applied to study electronic structure properties of one-dimensional (1D) hydrogen chain, diamond and LiH crystal with significantly reduced circuit depth while maintaining the same accuracy as the ADAPT-VQE algorithm. For example, numerical results demonstrate that the VQE-PT algorithms exhibit energy deviations of 0.1 kcal/mol for the LiH crystal with only one circuit parameter while the ADAPT-VQE requires $\sim$30 parameters to achieve the same accuracy. Meanwhile, the new algorithms can improve the accuracy of UCCSD and hardware efficient Ans\"atze with fixed circuit depth.}   

\section{Theory}

\subsection{Periodic electronic structure}
The second-quantized Hamiltonian in the Bloch molecular orbital representation is written as
\begin{equation}\label{eq:H}   
\hat{H} = E_{\mathrm{NN}} + \sum_{p\bk_p q\bk_q}^\prime \mathrm{h}^{p\bk_p}_{q\bk_q} \hat{T}^{p\bk_p}_{q\bk_q} + \frac{1}{2} \sum_{p\bk_p q\bk_q r\bk_r s\bk_s}^\prime \mathrm{V}^{p\bk_p q\bk_q}_{r\bk_r s\bk_s} \hat{T}^{p\bk_p q\bk_q}_{r\bk_r s\bk_s}.
\end{equation}
Here, single and double excitation operators are defined as
\begin{equation}\label{eq:pbc_t}
    \begin{split}
        \hat{T}^{p\bk_p}_{r\bk_r} &= \hat{a}_{p\bk_p}^\dag \hat{a}_{r\bk_r} \\   \hat{T}^{p\bk_p q\bk_q}_{r\bk_r s\bk_s} &= \hat{a}_{p\bk_p}^\dag \hat{a}_{q\bk_q}^\dag \hat{a}_{s\bk_s} \hat{a}_{r\bk_r}.
    \end{split}
\end{equation}
$a^\dagger$ and $a$ are creation and annihilation operators, respectively. The primed summation indicates that single and double excitation operators should satisfy the translational symmetry imposed by the periodic boundary condition, that is, the wave vectors $\bk$ in the excitation operators satisfy the conservation of momentum
\begin{equation}
  \sum_{p} \bk_{p}-\sum_{r} \bk_{r}=\boldsymbol{G}_m \label{conservation of momentum}
\end{equation}
with $\boldsymbol{G}_m$ being the reciprocal lattice vector. The one-electron matrix $\mathbf{h}$ consists of kinetic, ionic potential and external potential integrals; The two-electron matrix $\mathbf{V}$ includes two-electron Coulomb integrals; $E_{\mathrm{NN}}$ is the nuclear-nuclear repulsion energy. In the following, we will drop ``$\bk$'' index and use ``$p$'' to indicate ``$p\bk_p$'' for simplicity. 

\subsection{Adaptive variational quantum algorithms}
A quantum state can be generally represented by applying a unitary evolution operator $\hat{U}(\boldsymbol{\theta})$ onto an initial state $|\psi_0\rangle$:
\begin{equation}
|\Psi(\boldsymbol{\theta})\rangle=\hat{U}(\boldsymbol{\theta})|\psi_0\rangle. \label{eq:wavefunction}
\end{equation}
{ Given a parameterized ansatz $\hat{U}(\boldsymbol{\theta})$, the} variational parameters $\boldsymbol{\theta}$ are optimized based on the Rayleigh-Ritz variational principle
\begin{equation}   E=\min_{\boldsymbol{\theta}} \langle\Psi(\boldsymbol{\theta})|\hat{H}|\Psi(\boldsymbol{\theta})|\rangle \label{eq:energy}.
\end{equation}
Adaptive variational quantum algorithms, such as ADAPT-VQE~\cite{GriEcoBar19}, are able to build a compact ansatz circuit and in principle represent the exact wave function with arbitrarily long product of unitary exponentialized single and double excitation operators as
\begin{equation}
    |\Psi(\boldsymbol{\theta})\rangle=\prod^{\infty}_{k=1}{e^{\theta(k) \hat{\tau}(k) }|\psi_0\rangle}\label{eq:trotter_wavefunction}
\end{equation}
The anti-Hermitian operators are defined as
\begin{equation}\label{eq:tau}
    \hat{\tau}_\mu = \hat{T}_\mu - \hat{T}^\dagger_\mu,
\end{equation}
where $\hat{T}_\mu \in \{\hat{T}^p_q,\hat{T}^{pq}_{rs}\}$. { We can iteratively update the wave function with
\begin{equation} 
|\Psi(k)\rangle=e^{\theta_M^{(k)} \hat{\tau}_M^{(k)}} \cdots e^{\theta_1^{(k)} \hat{\tau}_1^{(k)}} |\Psi(k-1)\rangle \label{eq:k-th wavefunction}
\end{equation}
where $|\Psi(0)\rangle=|\psi_0\rangle$ is the initial state. In $k$-th iteration,
the operators $\{ \hat{\tau}_1^{(k)},\cdots,\hat{\tau}_M^{(k)} \}$ with the largest residual gradients
\begin{gather}
    g_\mu=\left. \frac{\partial \langle \Psi(k) | \hat{H} | \Psi(k) \rangle}{\partial \theta(k)} \right|_{\hat{\tau}(k)=\hat{\tau}_\mu,\theta(k)=0}
    \label{eq:g}
\end{gather}}
are used to update the wave function. The convergence criteria of the ADAPT-VQE algorithm is:
\begin{equation}
   |\boldsymbol{g}|_2=\sqrt{\sum_\mu {|g_\mu|^2}} < \epsilon \label{eq:conv}
\end{equation}
{
The algorithm flowchart of ADAPT-VQE is illustrated in Algorithm~\ref{alg:adapt}.
\begin{algorithm}[H]
\caption{The ADAPT-VQE algorithm for optimizing the wave function and the total energy.}

\leftline{\textbf{Input:} Reference state $|\psi_0\rangle$ and Hamiltonian $\hat{H}$.}

\leftline{\textbf{Output:} The total energy $E$ and the wave function $|\Psi\rangle$ of the target state.}

\begin{algorithmic}[1]

\STATE Prepare the initial wavefunction $|\Psi \rangle = |\Psi_0 \rangle$ in qubit representation.

\STATE Define the operator pool.

\STATE Initialize the operator sequence $\vec{\tau}=\{\}$ and parameters $\boldsymbol{\theta}=\{0\}$.

\WHILE {$|\boldsymbol{g}|_2 \geq \epsilon$}

\STATE Compute $g_\kappa$ using Eq.~\eqref{eq:g} for all $\tau_\kappa$ in the operator pool.

\STATE $\vec{\tau} \gets \{\vec{\tau}, \hat{\tau}_{1}^{(k)},\ldots,\hat{\tau}_{M}^{(k)}\} $ where $\{\hat{\tau}_l^{(k)}\}_{l=1}^{M} $ are $M$ operators with the largest absolute residual gradients and $\vec{\theta}=\{\vec{\theta},0,\ldots,0\}$.

\STATE Update the new wavefunction with Eq.~\eqref{eq:k-th wavefunction}.

\STATE Optimize parameters $\boldsymbol{\theta}$ with Eq.~\eqref{eq:energy}.

\ENDWHILE

\STATE Return $E=E(\boldsymbol{\theta}_{\mathrm{min}})$ and  $|\Psi\rangle = |\Psi(\boldsymbol{\theta}_{\mathrm{min}})\rangle$.

\end{algorithmic}
\label{alg:adapt}
\end{algorithm}
}
As discussed in Ref.~\citenum{LiuWanLi20}, the complex phase introduced by Bloch Hartree-Fock orbitals results in loss of accuracy in ADAPT-VQE using the operator pool defined with Eq.~\eqref{eq:tau}. Fan {\it et al.} suggested that additional anti-Hermitian operators
\begin{equation}
    \hat{\tau}_\mu = i(\hat{T}_\mu + \hat{T}^\dagger_\mu)
\end{equation}
should be introduced to recover the exact energy~\cite{FanLiuLi21}. Alternatively, one can transform Block Hartree-Fock orbitals at sampling $k$-points in a unit cell into orbitals at $\Gamma$-point of the corresponding supercell. This avoids the complex coefficients appearing in the Hamiltonian~\cite{LiuWanLi20}. A brief introduction of this scheme, named K2G, is described as following. Block Hartree-Fock orbitals can be reconstructed with
\begin{equation}
     \phi_{j\bk}(\br) = \sum_{\kappa_n} \tilde{C}_{\kappa_n j\bk } \chi_{\kappa_n}(\br),
\end{equation}
where the coefficients are
\begin{equation}
    \tilde{C}_{\kappa_n j\bk} = \frac{1}{\sqrt{N_L}} e^{i\bk \cdot \bR_n} C_{\kappa j}(\bk).
\end{equation}
$\chi_{\kappa_n}$ is the $\kappa$-th atomic orbital in $n$-th ``replica'' and $N_L$ is the number of ``replica''. $\bR_n$ is the translation vectors and $j\bk$ indicates $j$-th Hartree-Fock orbital at $\bk$ point. The Fock matrix with the supercell atomic orbital basis is expressed as
\begin{equation}
    F_{\kappa_m \lambda_n} = \sum_{j\bk} \tilde{C}_{\kappa_m j\bk} E_{j \bk} \tilde{C}_{\lambda_n j\bk}^\dag
\end{equation}
Diagonalizing this Fock matrix, we obtain the real orbitals expanded using the supercell atomic basis functions. { It is clear that after the K2G transformation, the VQE algorithm for periodic systems with multiple $k$ points and molecular systems is integrated into a unified framework. Therefore, the VQE-base perturbation theory introduced in the following is applicable to not only the periodic Hamiltonian of Eq.~\eqref{eq:H} but also the molecular Hamiltonian.}

\subsection{VQE-based perturbation theory}\label{ssec:PT}
{
To derive VQE-based perturbation theory, one should first determine the wave function ansatz from an ADAPT-VQE or UCCSD-VQE calculation. In case of ADAPT-VQE, a convergence threshold $\epsilon$ defined in Eq.~\eqref{eq:conv} is necessary to build the ansatz. In the UCCSD-VQE, the ansatz is well defined (see Eq.~\eqref{eq:uccsd}) so that one only requires to optimize the variational parameters to determine the ansatz. In this work, we use a large convergence threshold, e.g. $\epsilon=0.1$ for ADAPT-VQE, to generate a low-depth ansatz. After a VQE calculation, one can obtain the ansatz parameters $\boldsymbol{\theta}_\mathrm{min}$, which minimizes the energy functional of Eq.~\eqref{eq:energy}. The reference wave function, namely the zeroth-order wave function in the VQE-MRPT, is defined as
\begin{equation}\label{eq:ref_wf}
    |\Psi_0\rangle = \hat{U}_0 |\psi_0\rangle
\end{equation}
with $\hat{U}_0 = \hat{U}(\boldsymbol{\theta}_\mathrm{min})$. The reference state of Eq.~\eqref{eq:ref_wf} can in general be considered a multiconfigurational state and expanded as a linear combination of determinants. 
}

\subsubsection{Multireference perturbation theory}
Given a reference state of Eq.~\eqref{eq:ref_wf}, it is simple to construct a set of excited states
\begin{equation}\label{eq:ex}
   |\Psi_\mu \rangle =  \hat{\tau}_\mu |\Psi_0 \rangle,
\end{equation}
which are orthogonal to the reference wave function
\begin{equation}\label{eq:orth}
  \langle \Psi_0 | \Psi_\mu \rangle = 0
\end{equation}
if the wave function is real. { Note that instead of general excitation operators of Eq.~\eqref{eq:pbc_t} a set of anti-Hermitian excitation operators is used here to introduce orthogonality between the reference and excited states.} In the case of periodic electronic structure calculations with multiple $k$ points, one can apply the K2G transformation to Bloch Hartree-Fock orbitals in order to satisfy Eq.~\eqref{eq:orth}. The exact wave function can be approximated as
\begin{equation}\label{eq:wf}
    | \Psi \rangle \approx \sum_{\mu=0}^{N_\mu} d_\mu |\Psi_{\mu} \rangle.
\end{equation}
Projecting the Schr\"odinger equation onto the subspace $\{ |\Psi_{\mu}\rangle \}$, one can obtain the generalized eigenvalue equations
\begin{equation}
    \mathbf{Hd} = \mathbf{ESd}.
\end{equation}
Here, $H_{\mu\nu} = \langle \Psi_\mu | \hat{H} | \Psi_\nu \rangle $ and $S_{\mu\nu} = \langle \Psi_\mu | \Psi_\nu \rangle$. 

{ The VQE-PT method aims at treating strongly correlated problems so that the reference state $|\Psi_0\rangle$ derived from the the VQE should be able to capture most of the static correlation. In practice, a reference state is considered appropriate if the magnitude of the perturbation correction energy for this reference state is less than a specified threshold. Given a reasonable reference state,} it is able to rewrite the wave function of Eq.~\eqref{eq:wf} as
\begin{equation}\label{eq:wf_pt}
    \begin{split}
    | \Psi \rangle \approx | \Psi_0 \rangle + | \Psi^{(1)} \rangle \\
    | \Psi^{(1)} \rangle = \sum_{\mu=1}^{N_\mu} d_\mu |\Psi_{\mu} \rangle.
    \end{split}
\end{equation}
Here, $| \Psi^{(1)} \rangle$ is considered as the first-order correction to the reference state. Inserting Eq.~\eqref{eq:wf_pt} into the Schr\"odinger equation, 
\begin{equation}
    \hat{H}  (| \Psi_0 \rangle + | \Psi^{(1)} \rangle) \approx E (| \Psi_0 \rangle + | \Psi^{(1)} \rangle),
\end{equation}
one can formulate the perturbation correction to the total energy as
\begin{equation}
\begin{split}
    \delta E &= E - E_0 \\
    &=\sum_{\mu=1}^{N_\mu} d_\mu \langle \Psi_0 |\hat{H}| \Psi_\mu \rangle 
    \end{split}
\end{equation}
and the reference energy is defined as $E_0 = \langle \Psi_0|\hat{H}|\Psi_0 \rangle$. The coefficients $d_\mu$ are obtained by solving the linear algebra equation
\begin{equation}\label{eq:MRPT-Eigen}
    \sum_{\nu=1}^{N_\mu} (E_0 S_{\mu\nu} - H_{\mu\nu}) d_\nu = \langle \Psi_\mu |\hat{H}| \Psi_0 \rangle. 
\end{equation}
{ Note that this generalized eigenvalue problem is similar to those derived in quantum subspace expansion (QSE)~\cite{TakRubJia20} and quantum Krylov subspace methods~\cite{StaHuaEva20}. For QSE and Krylov subspace methods, the generalized eigenvalue problems are to diagonalize the Hamiltonian in a subspace. While, Eq.~\eqref{eq:MRPT-Eigen} is to calculate the coefficients for the perturbation correction in the total energy. In general, Eq.~\eqref{eq:MRPT-Eigen} is analogous to the generalized eigenvalue equation in QSE. However, in the QSE method, the reference wave function should be included in the subspace expansion and the excitation operators are unnecessary to be anit-Hermitian. }
{ This ``diagonalize-then-perturb'' procedure had also been used in the SCI- and DMRG-based perturbation theory to introduce a second-order perturbation correction~\cite{GarSceGin18,SonCheMa20}.}

Furthermore, one can perform a singular value decomposition (SVD) of the overlap matrix~\cite{AndMalRoo92,BurTho20}
\begin{equation}
    \mathbf{S = Q s V^\dagger}.
\end{equation}
{ Although the dimension of $\mathbf{S}$ scales in principle as $N^4$, we can prescreen $|\Psi_\mu\rangle$ with the coefficients $d_\mu$ as weights so that the dimension of SVD can be significantly reduced~\cite{ZhaLiuHof20}.} In order to remove the linear dependency of the wave functions $\{|\Psi_\mu\rangle\}_{\mu=1}^{N_\mu}$, the elements of $\mathbf{s}$ less than a specified threshold $\epsilon_r$ can be discarded. In this work, we set $\epsilon_r=10^{-10}$ in the following calculations. A set of orthogonal wave functions can be constructed as 
\begin{equation}
    |\tilde{\Psi}_\mu \rangle = \sum_{\nu=1}^{N_\mu} Q_{\mu\nu} |\Psi_\nu \rangle,
\end{equation}
where $\tilde{N}_\mu$ is the number of $|\tilde{\Psi}_\mu \rangle$, which is equal to or less than $N_\mu$. Consider a diagonal approximation to Eq.~\eqref{eq:MRPT-Eigen}, one can define perturbation correction to the total energy as
\begin{equation} \label{eq:mrpt-d}
    \delta E = \sum_{\mu=1}^{\tilde{N}_\mu} \frac{|\langle \Psi_0 |\hat{H}| \tilde{\Psi}_\mu \rangle|^2 }{E_0 - \tilde{E}_\mu}.
\end{equation}
Here, $ \tilde{E}_\mu = \langle \tilde{\Psi}_\mu | \hat{H} | \tilde{\Psi}_\mu \rangle$. 

\subsubsection{Similarity transformed perturbation theory}
The VQE energy can be reformulated as
\begin{equation}
    E_0 = \langle \Psi_0|\hat{H}|\Psi_0 \rangle =  \langle \psi_0|\bar{H}|\psi_0 \rangle
\end{equation}
where the similarity-transformed Hamiltonian { (which is also referred to the effective Hamiltonian)} is defined as
\begin{equation}
    \bar{H} = \hat{U}_0^\dagger \hat{H} \hat{U}_0.
\end{equation}
{ Consider the Hartree-Fock state $|\psi_0\rangle$ as the zeroth-order wave function, one can approximate the exact wave function as}
\begin{equation}\label{eq:wf_STPT}
    |\Psi\rangle \approx |\psi_0\rangle + \sum_{\lambda=1}^{N_\lambda} d_\lambda |\psi_\lambda \rangle  
\end{equation}
where $|\psi_\lambda\rangle$ consists of configuration state functions generated by applying the ``excitation'' operators in the similarity transformed Hamiltonian to the initial state $|\psi_0\rangle$ if one assumes $\langle \psi_0 |\bar{H}| \psi_\lambda \rangle \neq 0$. This scheme was first suggested in ref.~\citenum{RyaIZmGen21} to introduce a perturbation correction to iterative qubit coupled cluster. { In this work, we formulate the perturbation theory in the fermionic representation instead of the qubit representation used in ref.~\citenum{RyaIZmGen21}. In the following calculations, only single and double excitation configurations are included in order to reduce the computational cost.} We note that this will result in loss of accuracy when the effective Hamiltonian is composed of a large number of high-order ($>$2-order) many-body operators.

Similar to Eq.~\eqref{eq:wf_pt}, consider a linear combination of excitation configurations (the second term on the right hand side of Eq.~\eqref{eq:wf_STPT}) to be a perturbation correction to the wave function, the coefficients $\mathbf{d}$ are obtained by solving the equation
\begin{equation}\label{eq:vqe-stpt}
     E_0 d_\lambda  - \sum_{\kappa=1}^{N_\lambda} \bar{H}_{\lambda\kappa} d_\kappa = \langle \psi_\lambda |\bar{H}| \psi_0 \rangle. 
\end{equation}
Here, $\bar{H}_{\lambda\kappa} = \langle \psi_\lambda | \bar{H} | \psi_\kappa \rangle $. A diagonal correction to the total energy can be defined { by discarding the nondiagonal elements of Eq.~\eqref{eq:vqe-stpt}} as
\begin{equation} \label{eq:stpt-d}
    \delta E = \sum_{\lambda=1}^{N_\lambda} \frac{|\langle \psi_0 |\bar{H}| \psi_\lambda \rangle|^2 }{E_0 - E_\lambda}
\end{equation}
with $E_\lambda = \langle \psi_\lambda |\bar{H}| \psi_\lambda \rangle$. { In contrast with the VQE-MRPT approach, the VQE-STPT approach defines the perturbation correction with an effective Hamiltonian $\bar{H}$ and the Hartree-Fock reference state as the zeroth-order wave function. Therefore, the VQE-STPT only requires a quantum computer to determine the ansatz while the VQE-MRPT requires to measure high-order reduced density matrices on a quantum computer. However, the VQE-STPT needs to in principle include all excitation operators appearing in the effective Hamiltonian to calculate the perturbation correction. As such, the number of measurements in the VQE-STPT will increase much faster than that in the VQE-MRPT as the system size increases. On the other hand, the diagonal approximation to the VQE-STPT of Eq.~\eqref{eq:stpt-d} does not need to perform a SVD of excitation configuration basis functions since they are orthogonal.}

\section{Numerical results}
Numerical simulations are executed with Q2Chemistry~\cite{q2chemistry}. Mapping fermionic operators onto qubit operators is performed using OpenFermion~\cite{openfermion}. We use PYSCF~\cite{pyscf} to perform Hartree-Fock calculations and output one- and two-electron integrals. The variational parameters in the VQE ansatz circuit are optimized with the Broyden-Fletcher-Goldfarb-Shannon (BFGS) algorithm~\cite{scipy}. In the case of the UCCSD ansatz without Trotterization, gradients can be computed with the parameter shift~\cite{IzmLaYen21,KotAnaAsp21} or finite difference approach. All reference results are obtained by exactly diagonalizing the  model Hamiltonian. The GTH-SVZ basis set together with GTH-PADE pseudopotential is used for calculations of diamond and 1D hydrogen chain models with equivalent bond length. The GTH-DVZ basis set is used for LiH crystal and 1D hydrogen chain with alternative bond length. In the following, multireference and similarity transformation-based perturbation theories are abbreviated as ``MRPT'' and ``STPT'', respectively. 

\begin{figure}[!htb]   
	\centering
	\includegraphics[width=0.9\linewidth]{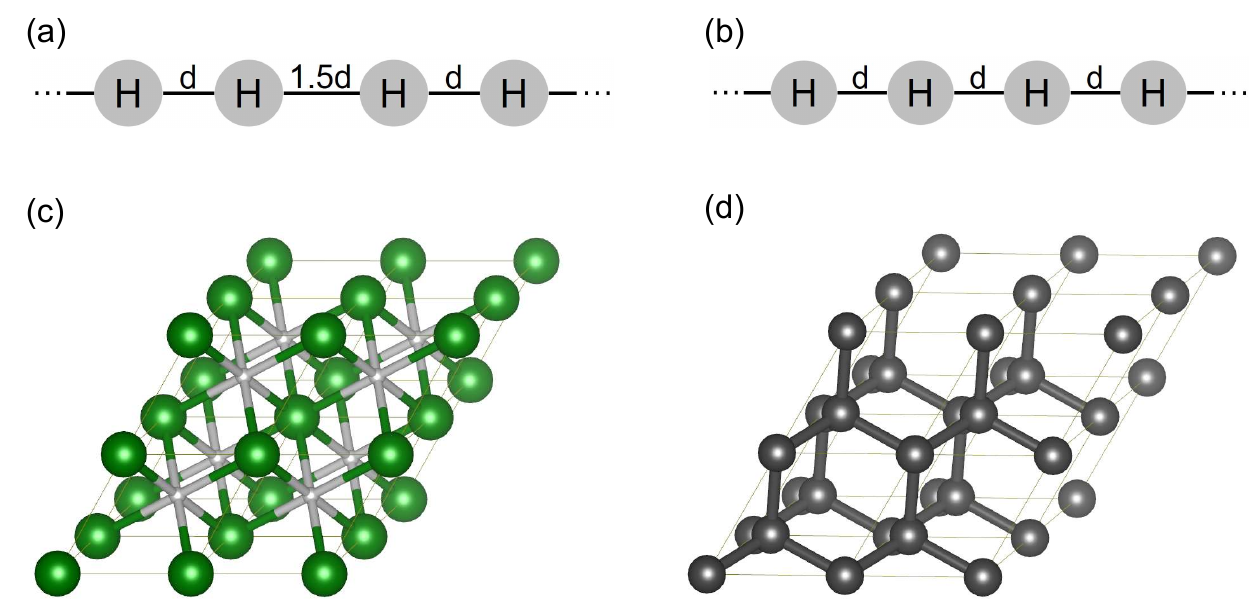}
	\caption{ Structures of (a) one-dimensional hydrogen chain with alternative bond lengths, (b) one-dimensional hydrogen chain with equivalent bond lengths, (c) LiH crystal with lattice constants $4.017$ \AA, and (d) diamond with lattice constants $3.567$ \AA .}
	\label{fig:structure}
\end{figure}

All calculations are performed with coarse $k$-point sampling grids (from $1\times1\times1$ to $1\times1\times4$). We note that the larger the number of $k$-point grids to sample the Brillouin zone is, the more accurate the simulation results are. The theoretical model with few k points exhibits significant finite-size effects and can not reproduce correct electronic structure properties in the thermodynamic limit. In this work, we aim to illustrate potential applications of our algorithms in materials science. The geometry structures of model systems used in this work are shown in figure~\ref{fig:structure}. In all cases, a unit cell consists of 2 atoms so that no more than 16 qubits are required in the VQE calculations. If not specified, spin-adapted fermionic excitation operators are used. 

\subsection{Accuracy and convergence}
The 1D hydrogen chain with two different H-H bond lengths in alternating positions along the chain, namely a Peierls-type distortion, is employed to assess the performance of different ADAPT-VQE-PT approaches. As shown in figure~\ref{fig:structure}(a), the short H-H bond length is set to be $d=1.5$ \AA \, and the long one is set to be $1.5d$. As stated above, the reference state $|\Psi_0\rangle$ is built using the ADAPT-VQE, in which the ansatz circuit is updated using five operators with the largest residual gradients in each iteration. The suffix ``-D'' indicates that a diagonal approximation of the perturbation correction (Eq.~\eqref{eq:mrpt-d} and ~\eqref{eq:stpt-d}) is used. Here, all calculations are run using $1\times1\times4$ $k$-points with  
four $k$-points sampled along the hydrogen chain and one $k$-point is
sampled along other two orthogonal directions. The lowest two Hartree-Fock orbitals are included in the active space and thus 16 qubits are used in the VQE calculations. 

\begin{figure}[!htb]   
	\centering
	\includegraphics[width=0.5\linewidth]{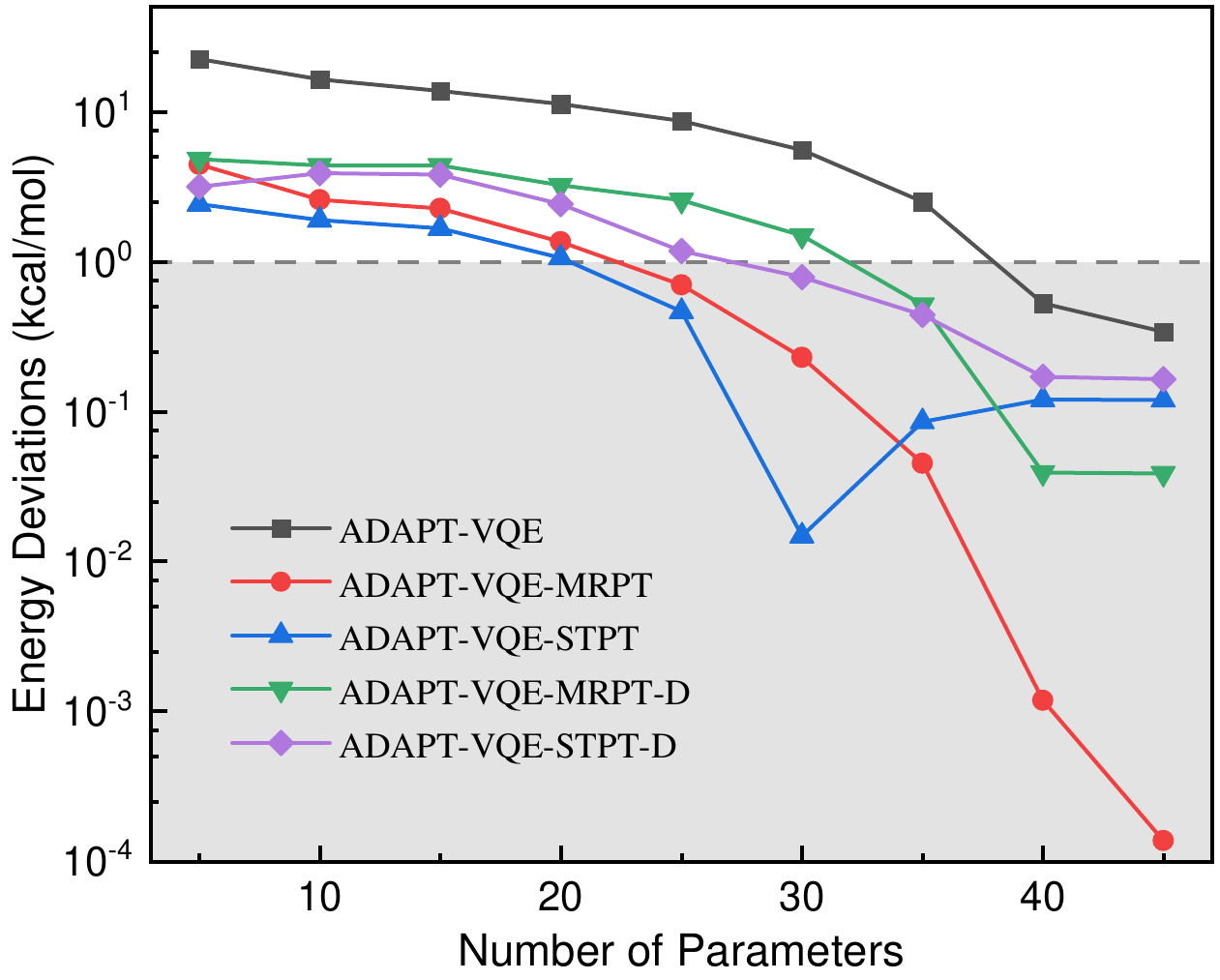}
	\caption{Energy deviations ($\Delta E = |E-E_\mathrm{exact}|$ in kcal/mol) as a function of the number of variational parameters in the ADAPT-VQE and ADAPT-VQE-PT approaches for one-dimensional hydrogen chain.}
	\label{fig:convergence}
\end{figure}

Figure~\ref{fig:convergence} shows derivations in the total energy as a function of the number of variational parameters ($N_p$) used in the ADAPT-VQE calculations. Results of ADAPT-VQE, ADAPT-VQE-MRPT, ADAPT-VQE-STPT, ADAPT-VQE-MRPT-D, and ADAPT-VQE-STPT-D are exhibited for comparison. ADAPT-VQE is able to achieve chemical accuracy (with the energy deviation less than 1 kcal/mol) using an ansatz state consisting of 40 variational parameters. In contrast, energy deviations of ADAPT-VQE-MRPT and ADAPT-VQE-STPT using 25 variational parameters are as small as 0.70 and 0.47 kcal/mol, respectively. Although energy deviations of ADAPT-VQE-MRPT and ADAPT-VQE-STPT are much smaller than those of ADAPT-VQE, they decrease at a similar rate before $N_p = 20$. After that, ADAPT-VQE-MRPT converges fast to a very high accuracy as $N_p$ increases ($\Delta E \sim 10^{-4}$ when $N_p = 45$). While, energy deviations of ADAPT-VQE-STPT stop decreasing in a monotonic manner after $N_p = 30$. This results from the fact that only single and double excitation configurations are employed in the ADAPT-VQE-STPT approach. As $N_p$ increases, the perturbation corrections from higher excitation configurations may become important if the weights of corresponding excitation operators in the effective Hamiltonian are large enough. Due to the diagonal approximation introduced in the ADAPT-VQE-MRPT-D and ADAPT-VQE-STPT-D, they converge in the total energy more slowly than ADAPT-VQE-MRPT and ADAPT-VQE-STPT.   

\subsection{ LiH crystal and diamond}
We apply different perturbative VQE methods, including ADAPT-VQE-MRPT, ADAPT-VQE-STPT, ADAPT-VQE-MRPT-D and ADAPT-VQE-STPT-D, to study ground-state energy curves of LiH crystal and diamond over a range of lattice parameters. Here, the equilibrium lattice constant is scaled by a factor ranging from 0.8 to 1.2. The reference wave function is generated using the ADAPT-VQE method with a convergence threshold of $\epsilon=0.1$. In the ADAPT-VQE calculations, only a single parameter is updated at a time. For diamond, only $\Gamma$-point is considered so that 16 spin orbitals are used in the VQE calculations. For LiH crystal, $1\times2\times2$ $k$-point grids are used to sample in the Brillouin zone. In addition, we frozen the lowest Hartree-Fock orbital and include another occupied orbital and the lowest unoccupied orbital in the active space.  

\begin{figure}[!htb]   
	\centering
	\includegraphics[width=0.9\linewidth]{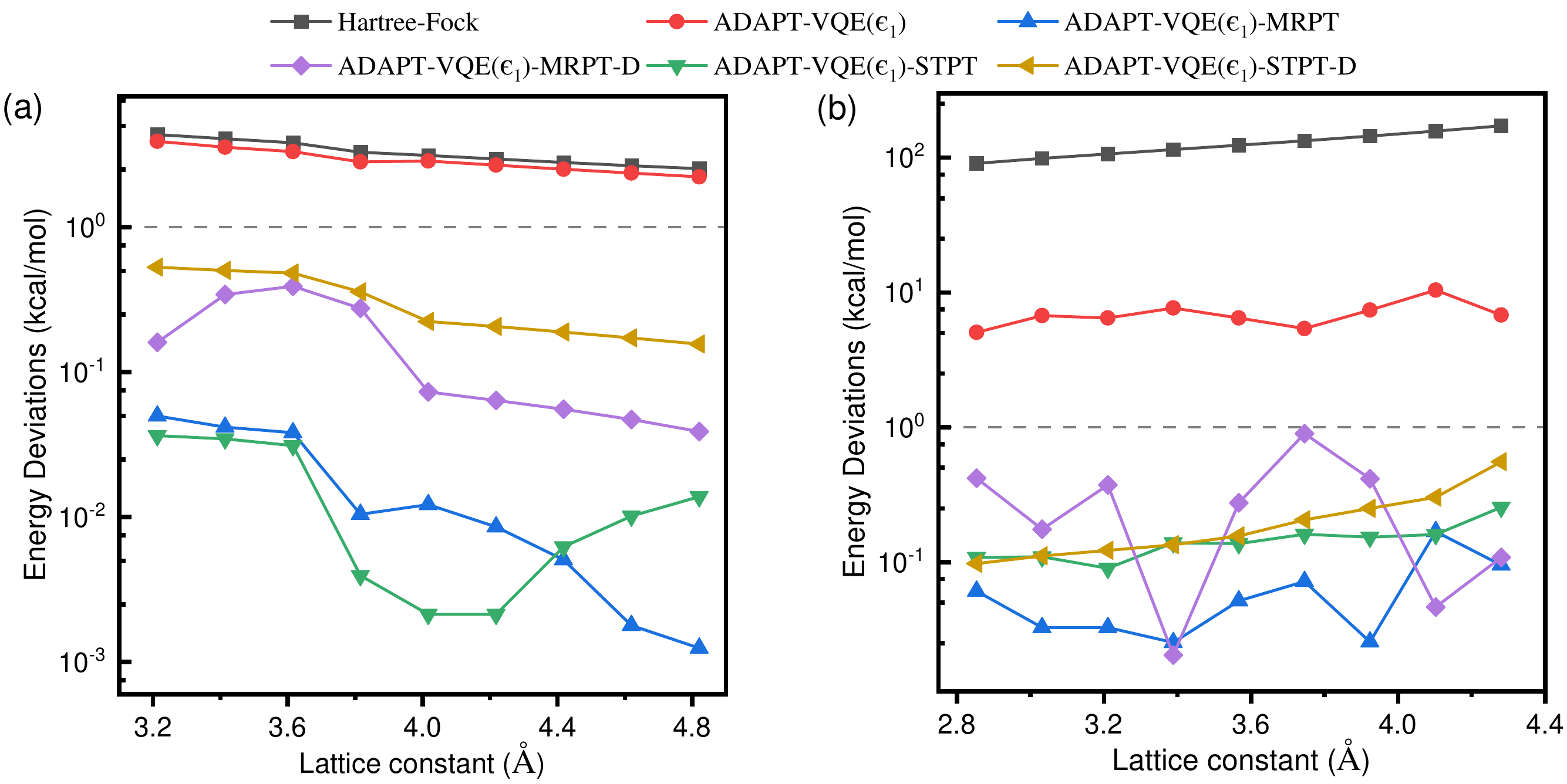}
	\caption{Ground-state energy deviations (in kcal/mol) of (a) LiH crystal and (b) diamond as a function of lattice constant. ``ADAPT-VQE(0.1)'' indicates a ADAPT-VQE calculation with $\epsilon=0.1$.}
	\label{fig:diamond}
\end{figure}

Figure~\ref{fig:diamond}(a) shows deviations in the ground-state energy of LiH crystal as a function of lattice constant. The results computed using Hartree-Fock and ADAPT-VQE($\epsilon_1$) are also shown for comparison. The errors in the total energy for Hartree-Fock are quite small for this model system, with a maximal error of 4.35 kcal/mol, since 20\% bond length change does not induce significant degenerate states in this model. The total energies computed with ADAPT-VQE($\epsilon_1$) are quite close to corresponding Hartree-Fock results because the ansatz state with a single parameter is able to converge to the specified threshold of $\epsilon=10^{-1}$. Both ADAPT-VQE($\epsilon_1$)-MRPT and ADAPT-VQE($\epsilon_1$)-STPT are able to reproduce accurate results with energy deviations less than 0.1 kcal/mol. ADAPT-VQE($\epsilon_1$)-MRPT-D and ADAPT-VQE($\epsilon_1$)-STPT-D are also able to achieve chemical accuracy for this model.

In the case of diamond, the correlation energy is much larger than that of LiH crystal while it still change gently over a range of lattice constant. In contrast with Hartree-Fock, ADAPT-VQE($\epsilon_1$) is able to reduce the errors in the total energy by one order of magnitude, with a maximal one of 10.38 kcal/mol. Correspondingly, the reference wave function consists of tens of variational parameters. The overall energy deviations of ADAPT-VQE($\epsilon_1$)-MRPT are less than 0.1 kcal/mol except for one of 0.17 kcal/mol at lattice constant of 4.1 \AA. The performance of ADAPT-VQE($\epsilon_1$)-STPT is also very satisfying, with energy deviations ranging from 0.09-0.25 kcal/mol. Energy deviations of ADAPT-VQE($\epsilon_1$)-MRPT-D exhibits a more intense fluctuation than those of ADAPT-VQE($\epsilon_1$)-STPT-D, which monotonically increases as the lattice constant becomes large.

Table~\ref{table:Np} shows the number of variational parameters and energy derivations over a range of lattice constant for ADAPT-VQE($\epsilon_2$), ADAPT-VQE($\epsilon_3$), and ADAPT-VQE($\epsilon_1$)-MRPT. Here, different lattice parameters is labeled by the scale factor $s$. In the context of ADAPT-VQE, the number of variational parameters that determines the circuit depth depends heavily on the convergence criteria. For diamond, the errors in the total energy of ADAPT-VQE($\epsilon_1$)-MRPT are even smaller than those of ADAPT-VQE($\epsilon_2$) and ADAPT-VQE($\epsilon_3$). Meanwhile, the number of variational parameters used in ADAPT-VQE($\epsilon_1$)-MRPT is also smaller than that in ADAPT-VQE($\epsilon_2$) and ADAPT-VQE($\epsilon_3$). For example, in the case of $s=1.2$, the number of parameters used in ADAPT-VQE($\epsilon_3$) is $\sim$8 times more than that in ADAPT-VQE($\epsilon_1$)-MRPT. For LiH, ADAPT-VQE($\epsilon_1$)-MRPT yields much more accurate energies than ADAPT-VQE($\epsilon_2$) while it performs similarly with ADAPT-VQE($\epsilon_3$). As mentioned above, ADAPT-VQE($\epsilon_1$)-MRPT requires only one single parameter to generate the reference wave function so that it can achieve high accuracy with a shallow circuit. Meanwhile, the number of measurements may increase due to additional calculations of matrix elements of the Hamiltonian. 

\begin{table}[!htb]
    \centering
    \caption{The number of variational parameters ($N_p$) and energy deviations ($\delta E$ in kcal/mol) for different ADAPT-VQE and ADAPT-VQE-MRPT approaches. ``VQE($\epsilon_k$)'' is an abbreviation of an ADAPT-VQE calculation with $\epsilon=10^{-k}$. $s$ indicates the scale factor of lattice constant.}
    \begin{tabular}{cccccccccccccc}
        \hline
        &\multicolumn{6}{c}{Diamond} && \multicolumn{6}{c}{LiH}\\
        &\multicolumn{2}{c}{VQE($\epsilon_1$)-PT}&\multicolumn{2}{c}{VQE($\epsilon_2$)} & \multicolumn{2}{c}{VQE($\epsilon_3$)}  &&\multicolumn{2}{c}{VQE($\epsilon_1$)-PT}&\multicolumn{2}{c}{VQE($\epsilon_2$)} & \multicolumn{2}{c}{VQE($\epsilon_3$)}  \\
        \cline{2-3}\cline{4-5}\cline{6-7}\cline{9-10}\cline{11-12}\cline{13-14}
        $s$ & $N_p$ & $\delta E$&  $N_p$ & $\delta E$& $N_p$ & $\delta E$&& $N_p$ & $\delta E$& $N_p$ & $\delta E$& $N_p$ & $\delta E$\\  0.8&28&0.06&80&0.12&102&0.10&&1&0.05&32&0.09&50&0.00\\    0.9&40&0.03&87&0.15&103&0.12&&1&0.04&23&0.16&34&0.00\\ 
        1.0&24&0.05&53&0.19&100&0.12&& 1&0.01&16&0.22&28&0.03\\ 
        1.1&17&0.03&57&0.29&108&0.13&& 1&0.01&15&0.22&28&0.03\\ 
        1.2&16&0.10&60&0.60&129&0.23&& 1&0.00&15&0.20&28&0.04\\
        \hline
    \end{tabular}
    \label{table:Np}
\end{table}

\subsection{QEB-ADAPT-VQE reference wave function}\label{ssec:QEB}
Except for chemically inspired reference wave functions, hardware-efficient wave function ansatz, such as qubit excitation-based ansatz~\cite{YorArmBar21} and block-entangled circuit ansatz, are also widely used in the context of quantum computing. Here, we take QEB-ADAPT-VQE as an example to illustrate hardware-efficient ansatz-based perturbation theory. The QEB operator pool, including $\{Q_p^\dag Q_q - Q_q^\dag Q_p, Q_p^\dag Q_q^\dag Q_r Q_s - Q_s^\dag Q_r^\dag Q_q Q_p\}$, is defined with qubit creation operator $Q^\dagger=(X-iY)/2$ and annihilation operator $Q=(X+iY)/2$. Here, $X$ and $Y$ are Pauli-X and Pauli-Y operators. In order to converge the ansatz state to the correct spin symmetry, a penalty function is often added to the Hamiltonian
\begin{equation}
    \hat{H} = \hat{H} + \frac{\alpha}{2} [\hat{S}^2-S(S+1)]^2,
\end{equation}
where $\hat{S}$ is the spin operator and $S$ is the spin multiplicity. In the following calculations, $\alpha$ is set to 0.5 Hartree. In contrast to the spin-adapted fermionic operator pool, we set $\epsilon=0.05$ since QEB-ADAPT-VQE need a tighter convergence criteria to restore the symmetry. All calculations are run using $1\times1\times4$ $k$-point grids.

\begin{figure}[!htb]   
	\centering
	\includegraphics[width=0.95\linewidth]{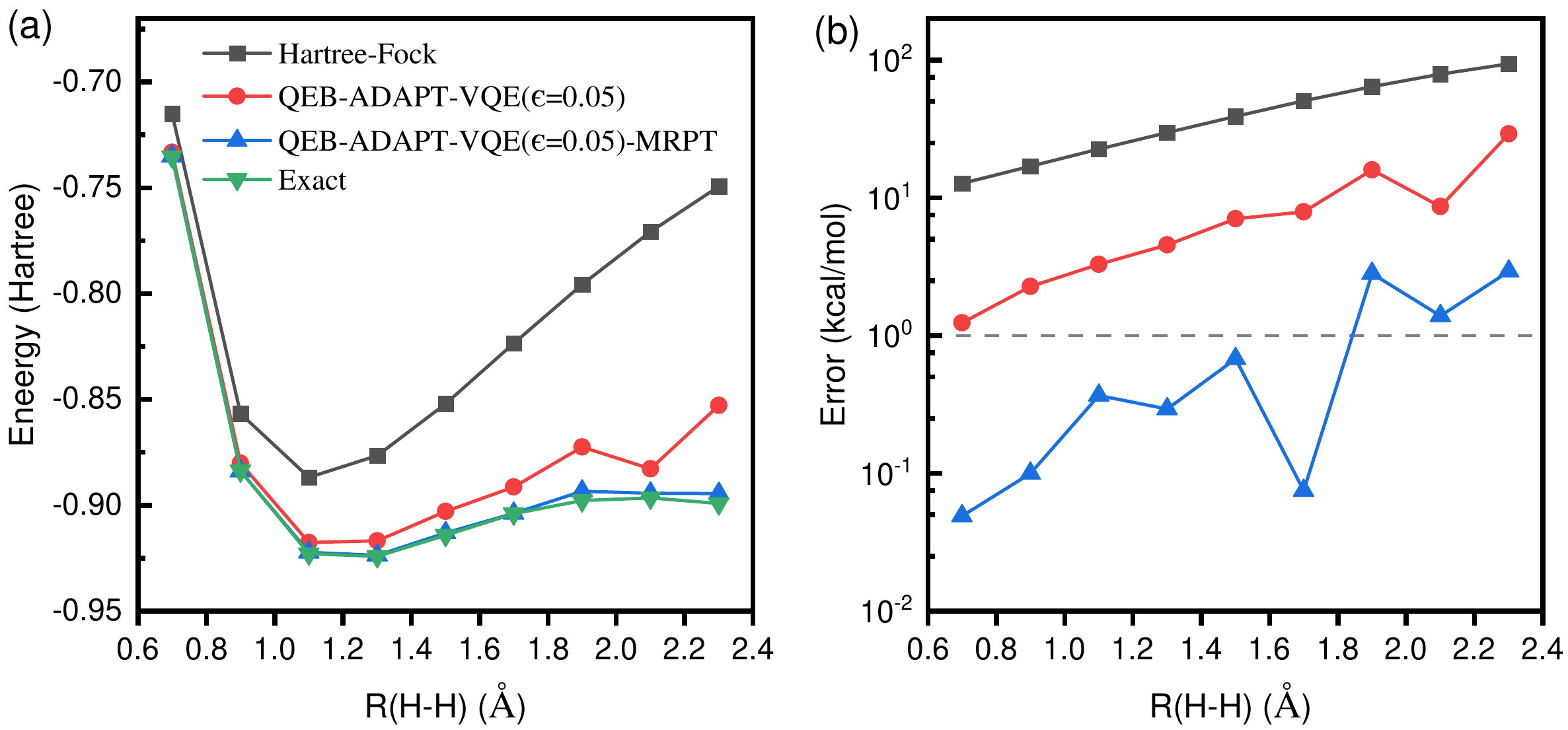}
	\caption{Ground-state energies (in Hartree) (a) and energy deviations (in kcal/mol) (b) computed with Hartree-Fock, QEB-ADAPT-VQE and QEB-ADAPT-VQE-MRPT as the function of the H-H bond length for one-dimensional hydrogen chain with equivalent bond lengths.}
	\label{fig:qeb}
\end{figure}

Figure~\ref{fig:qeb} shows ground-state energy curves and deviations as a function of the H-H bond length for the 1D hydrogen chain with equivalent bond length, which exhibits strong correlation effect as the H-H bond elongates. Due to the loose convergence criteria, the energy curves have small fluctuations as discussed in previous works~\cite{GriEcoBar19}. The errors in the total energy of QEB-ADAPT-VQE($\epsilon=0.05$) quickly increase as the bond length becomes large. Accordingly, deviations in the total energy of QEB-ADAPT-VQE($\epsilon=0.05$)-MRPT increase as those of QEB-ADAPT-VQE($\epsilon=0.05$) increase. The largest energy deviations of QEB-ADAPT-VQE($\epsilon=0.05$) and QEB-ADAPT-VQE($\epsilon=0.05$)-MRPT are as large as 29.08 and 2.93 kcal/mol at the H-H bond length of 2.3 \AA, respectively. Overall, the energy deviation of QEB-ADAPT-VQE-MRPT is positively related to that of QEB-ADAPT-VQE, that is the performance of the VQE-PT method depends on the reference wave function. 

\subsection{UCCSD reference wave function} \label{ssec:uccsd}
UCCSD has been successfully applied to study electronic structure properties within the VQE framework~\cite{SheZhaZha17,RomBabMcC18}. The UCCSD ansatz is written as
\begin{equation}\label{eq:uccsd}
|\Psi(\boldsymbol{\theta})\rangle=e^{ \sum_{\mu} \theta_\mu \hat{\tau}_\mu }|\psi_0\rangle, 
\end{equation}
where $\hat{\tau}_\mu \in \{\hat{\tau}_i^a,\hat{\tau}_{ij}^{ab}\}$, namely single and double excitations from occupied to unoccupied space. UCCSD is more robust than traditional coupled cluster with single and double excitations as demonstrated in previous studies~\cite{CooKno10,LeeHugHea19}. Here, the UCCSD ansatz is used as a reference wave function in the VQE-PT method. We name this approach as UCCSD-VQE-MRPT. In principle, each optimization should be carried out tens or even hundreds times with different initial guesses in order to approximate the true minimum. In this work, the parameters of the UCCSD ansatz are optimized only once with their initial values all set to zero. The optimization procedure finishes when the norm of gradients is less than $10^{-4}$. 

\begin{figure}[!htb]   
	\centering
	\includegraphics[width=1.0\linewidth]{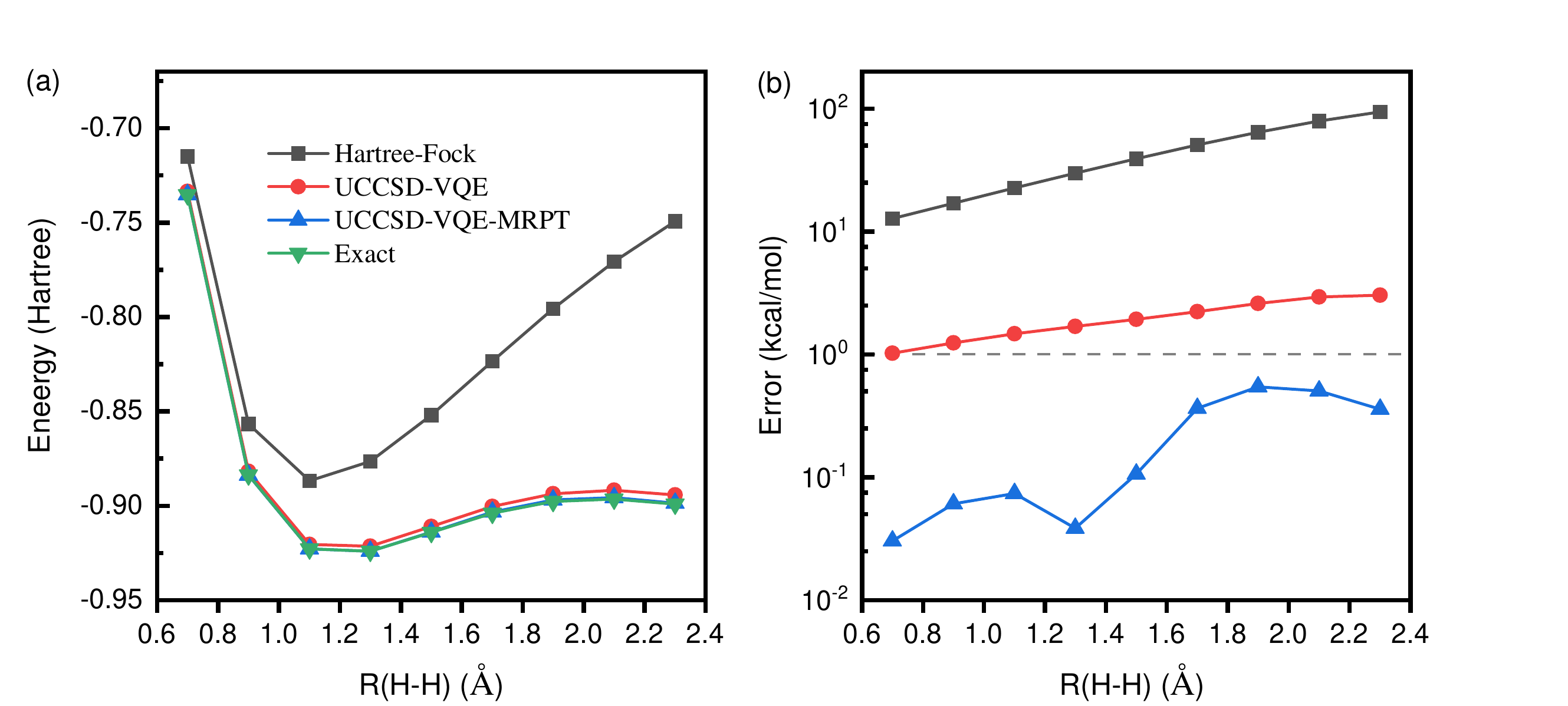}
	\caption{Ground-state energies (in Hartree) (a) and energy deviations (in kcal/mol) (b) computed with Hartree-Fock, UCCSD-VQE and UCCSD-VQE-MRPT for one-dimensional hydrogen chain with equivalent bond length.}
	\label{fig:uccsd}
\end{figure}

In figure~\ref{fig:uccsd}, we assess the performance of UCCSD-VQE-MRPT on the ground state of 1D hydrogen chain as introduced in section~\ref{ssec:QEB}. It is clear that the performance of Hartree-Fock and UCCSD-VQE deteriorates as the H-H bond length increases. For example, the energy deviation of UCCSD-VQE increases from 1.0 kcal/mol to 3.0 kcal/mol as the H-H bond length increases from 0.7 to 2.3 \AA. This can be foreseen since UCCSD-VQE is difficult to describe strongly correlated systems. In contrast, the maximal error of UCCSD-VQE-MRPT is 0.55 kcal/mol at the H-H bond length of 1.9 \AA. Therefore, UCCSD-VQE-MRPT is an appealing scheme to improve the accuracy of UCCSD-VQE.

\section{Conclusions}
Accurate prediction of electronic structure properties requires a reasonable description of both static and dynamic electron correlation. In this work, we present two schemes to implement perturbative variational quantum algorithms for material simulations by integrating adaptive variational quantum eigensolver and perturbation theory to quantify electron correlation. In the first scheme, we take the VQE ansatz state as the reference wave function and formulate VQE-based multireference perturbation theory in an excitation configuration space consisting of a set of wave functions orthogonal to the reference wave function using anti-Hermitian operator projecting. In the second scheme, we use the VQE ansatz circuit that represents a unitary transformation to perform similarity transformation of the Hamiltonian and then formulate the perturbation correction based on this effective Hamiltonian and corresponding excitation configurations. In these two schemes, the adaptive VQE ansatz is built using a loose convergence criteria so that the ansatz circuit is shallow enough while it can capture most of the correlation energy. The rest correlation energy, often related to the dynamic electron correlation, is restored by the perturbation theory. Applying these two adaptive VQE-PT algorithms to LiH crystal and diamond exhibits an accurate prediction of ground-state energies over a range of lattice constant.

Recently, a variety of ansatz have been proposed to achieve delicate balance between accuracy and circuit depth in the context of quantum computing. Hence, we also assess the performance of VQE-MRPT using different reference wave functions generated from various ansatz, including the UCCSD and QEB-ADAPT ansatz. In the former case, the VQE-MRPT scheme provides a promising tool to improve the accuracy of the UCCSD ansatz at a moderate computational cost. On the other hand, the UCCSD ansatz can be also systematically improved by including higher-order excitation operators at the expense of rapidly increasing of the number of variational parameters. This most probably comes with the notorious problem of high-dimensional nonlinear optimization in the VQE. In the latter case, due to the lack of physical symmetry, hardware-efficient ansatz converges much more slowly than chemically inspired ansatz, such as spin-adapted fermionic excitation-based ansatz, within the framework of adaptive variational quantum algorithms. The VQE-PT algorithm is a feasible way to avoid tedious iterative convergence processes. In addition, the accuracy of the VQE using a block-entangled ansatz depends on the expressive power of the ansatz circuit. The VQE-MRPT scheme can also efficiently improve the accuracy of hardware-efficient ansatz while maintaining a shallow circuit depth. Consider that the reference wave function is unnecessary to be a globally minimized VQE ansatz state, the potential ``barren plateaus'' problem in the VQE optimization may be no longer a problem in the VQE-PT method. 

The major computational cost of the VQE-PT method results from computing the matrix elements of the Hamiltonian in the perturbative space. In the VQE-MRPT scheme, the computational complexity scales as $N_H N_{\mu}^2$, with $N_H$ being the number of terms in the Hamiltonian. The number of single and double excitation anti-Hermitian operators $N_\mu$ in principle scales as $N^4$, with $N$ being the number of qubits. However, it is possible to define the importance of different excited states in the perturbative subspace by precalculating the coupling $\langle \Psi_\mu |\hat{H}|\Psi_0 \rangle$ and thus discard these ``unimportant' wave functions. In addition, in the VQE-MRPT-D scheme, the computational complexity scales as $N_H N_{\mu}$ if $|\tilde{\Psi}_\mu\rangle$ can be efficiently prepared on a quantum computer~\cite{KirMotMez23}. However, this may result in loss of the computational accuracy due to the diagonal approximation. In the VQE-STPT scheme, the computational complexity scales as $N_H N_{\lambda}^2$, with $N_H$ being the number of terms in the effective Hamiltonian. As such, its computational cost depends heavily on the similarity transformation. A complex ansatz circuit will result in an effective Hamiltonian consisting of a large number of terms and as a consequence a perturbative subspace consisting of single and double excitation configurations used in this work is most probably insufficient to estimate the perturbation correction. While, including all excitation configurations generated from applying the operators in the effective Hamiltonian to the Hartree-Fock wave function is computationally demanding except for some simple cases. Therefore, { in order to apply the VQE-PT algorithm to complex systems,} it is interesting to explore appropriate schemes to reduce the computational complexity without loss of accuracy within the VQE-PT framework in the future work. 

{Although the perturbative quantum algorithms are originally formulated based on the VQE in this work, one can also use the adiabatic evolution approach to prepare a multiconfigurational reference state~\cite{DuXuPen10}. The ADAPT-VQE-PT algorithms proposed in this work require remarkably a much shallower circuit to achieve the same accuracy as the ADAPT-VQE.  As such, our algorithm can mitigate error  from state preparation while the measurement error may become larger since the VQE-based perturbation theory requires at least fourth-order reduced density matrices measured on a quantum computer. While, the VQE-PT algorithms presented in this work can be easily extended to active space methods, in which the VQE calculation is executed in a very small active space and correspondingly the number of measurements for extracting reduced density matrices is also significantly reduced.}

\section{Acknowledgments}
J. L. Thanks Yang Guo for helpful discussions. This work was supported by Innovation Program for Quantum Science and Technology (2021ZD0303306), the National Natural Science Foundation of China (22073086, 22288201), Anhui Initiative in Quantum Information Technologies (AHY090400), and the Fundamental Research Funds for the Central Universities (WK2060000018).

\section{Conflicts of interest}
There are no conflicts of interest to declare.

\bibliography{qc}

\end{document}